\documentclass[10pt,aps,prb,final,twocolumn,amsmath,amssymb,superscriptaddress,longbibliography]{revtex4-1}

\usepackage[pdftex, colorlinks=true, linkcolor=blue, citecolor=blue, urlcolor=blue]{hyperref}
\usepackage{graphicx}
\usepackage{aurical}
\usepackage[T1]{fontenc}
\usepackage{units}

\DeclareGraphicsExtensions{.png,.pdf}

\newcommand{\fwidth}{0.85\linewidth}
\newcommand{\br}{\mathbf{r}}
\newcommand{\dif}{\mathrm{d}}
\newcommand{\EF}{\varepsilon_\mathrm{F}}
\newcommand{\bepsilon}{\bar{\varepsilon}}
\newcommand{\bl}{\bar{l}}

\newcommand{\ipcms}{Universit\'e de Strasbourg, CNRS, Institut de Physique et de Chimie des Mat{\'e}riaux de Strasbourg, UMR 7504, F-67000 Strasbourg, France}
\newcommand{\pullman}{Department of Physics and Astronomy, PO Box 642814, 
	Washington State University, Pullman, Washington 99164-2814, USA}

\begin{document}
	
\title{Partial local density of states from scanning gate microscopy}
\author{Ousmane Ly}
\affiliation{\ipcms}
\author{Rodolfo A.\ Jalabert}
\affiliation{\ipcms}
\author{Steven Tomsovic}
\affiliation{\pullman}
\author{Dietmar Weinmann}
\affiliation{\ipcms}

\begin{abstract}
Scanning gate microscopy (SGM) images from measurements made in the vicinity of 
quantum point contacts (QPC) were originally interpreted in terms of current flow.  Some recent work has analytically connected the local density of states to conductance changes in cases of perfect transmission, and at least qualitatively for a broader range of circumstances.  In the present paper, we show analytically that in any time-reversal invariant system there are important deviations that are highly sensitive to imperfect transmission.  Nevertheless, the unperturbed \textit{partial} local density of states can be extracted from a weakly invasive scanning gate microscopy experiment, provided the quantum point contact is tuned anywhere on a conductance plateau.  A perturbative treatment in the reflection coefficient shows just how sensitive this correspondence is to the departure from the quantized conductance value and reveals the necessity of local averaging over the tip position.  It is also shown that the quality of the extracted partial local density of states decreases with increasing tip radius.
\end{abstract}

\maketitle

\section{Introduction}
\label{sec:introduction}
Since its development 20 years ago,\cite{eriksson1996} 
scanning gate microscopy (SGM) has revealed fascinating phenomena 
in transport processes and has been considered as a powerful tool to probe 
local properties.\cite{topinka2000,topinka2002} In this technique the 
conductance of an electronic device is measured while the tip of an 
atomic force microscope (AFM) is scanned above its surface.  
The AFM tip acts as a movable gate  that scatters the electrons 
leading to a spatially dependent modulation of the conductance.\cite{sellier2011}
  
One of the most investigated  nanostructures is the quantum point contact 
(QPC)\cite{wees1988,wharam1988} defined in a two dimensional electron gas (2DEG).
When the tip is raster-scanned over the surface of the system, 
electrons are back-scattered to the QPC giving rise to a conductance map that
exhibits a branched pattern.   In the case of a QPC opening into an unconstrained 
2DEG these patterns have been interpreted as a signature of the electron flow in 
the disordered potential resulting from the 
ionized donor atoms.\cite{topinka2001,jura2007}
Thus, a link is presumed to exist between SGM measurements and local properties 
(local densities of states [LDOS] and current densities) of the unperturbed devices.

Typically, the tip voltages used to study QPC setups operating in the 
regime of conductance quantization are strong enough to create 
a large depletion disk (much bigger than the Fermi wavelength) in the 2DEG underneath 
the tip. The connection with local properties has been argued to concern the classical turning point of the electron 
trajectories with the Fermi energy that leave the QPC and encounter the 
tip potential.\cite{heller2003}  

In order to address this problem, the paradigmatic case of a QPC perturbed by a weakly invasive tip has been considered in the linear \cite{jalabert2010,gorini2013} and 
non-linear \cite{gorini2014} regimes (in source-drain bias voltage).
In particular, in the regime of conductance quantization of clean 2DEGs, 
spatial and time-reversal symmetries have been shown to play a key role in 
establishing a correspondence of the SGM response with the LDOS and the 
current density on both sides of the QPC.

The SGM technique has also been used to study systems with a variety of 
electronic confinements, including open quantum dots 
\cite{crook2003,burke2010,kozikov2013,steinacher2015,poeltl2016,kozikov2016}
and Aharonov-Bohm rings built in high-mobility semiconductor 
heterostructures, \cite{hackens2006,martins2007,pala2008,pala2009} as well as 
carbon nanotubes \cite{woodside2002} and graphene-based 
microstructures. \cite{schnez2011,cabosart2014} 
For systems with sufficient electronic confinement charging effects are relevant, and for very 
small quantum dots a biased SGM tip mainly acts as a gate that modifies the 
number of electrons in the dot and affects the conductance via the 
Coulomb-blockade phenomenon. 
\cite{pioda2004,fallahi2005,woodside2002,bleszynski2007,schnez2011} 

For relatively large and open quantum dots, the charging effects are not 
crucial and, as in the case of QPC setups, the connection between the SGM 
measurements and local properties has been pursued.  In these systems, 
qualitative similarity between conductance changes and LDOS has been 
noted whenever the LDOS exhibits some localized structure. For instance, 
minima of the SGM response appear  where the LDOS vanishes.\cite{pala2008,pala2009}  
Furthermore, numerical simulations for rectangular resonant 
cavities \cite{kolasinski2013} indicated that the conductance terms
derived in Ref.\ \onlinecite{jalabert2010} are correlated with the 
LDOS when the Fermi energy is close to a resonance with a cavity state.  
For one-dimensional systems, a perturbative approach has revealed that 
the first-order conductance change in the presence of a $\delta$-tip is 
related to the Hilbert transform of the LDOS. \cite{gasparian1996,pala2008}  

It is important to note that electronic confinement is associated with a 
change in the interpretation of SGM maps with respect to the case of a QPC.  
Specifically tailored experiments have shown the need of such a change of 
interpretation when the QPC setup is modified by electronic 
confinement guiding the electron transport.\cite{steinacher2016,kozikov2016}    
The need of different interpretations for setups with and without electronic 
confinement can be traced, in the case of weakly invasive probes, to special 
features of conductance quantization characterizing QPCs in the absence of 
confinement, where the transmission channels are either completely open or 
closed.\cite{jalabert2010}

The issue of whether the transmission channels are completely open 
(and otherwise completely closed), i.e.~the perfect transmission case, 
turns out to play a crucial role in the interpretation of measurements 
and their relationships to local properties.  It has been shown that in the 
case of perfect transmission, the second order conductance change is the first 
non-vanishing term in a perturbation series\cite{jalabert2010} and it is 
proportional to the square of the LDOS.\cite{gorini2013} 
However, the analytic relationship between conductance changes and local 
properties becomes more complicated for imperfect transmission.

In this paper analytical and numerical approaches are developed to study the 
connection between SGM measurements and local properties for the case 
of a QPC in which the tip potentials can be perturbative or non-perturbative, 
local or extended, etc.  In addition, cases where the 2DEG surrounding the QPC 
can be disordered or clean are treated.  First, in the perturbative regime and 
on a perfect conductance plateau (i.e.~at perfect transmission), 
the SGM on one side of the QPC is unambiguously related to the {\it partial}
LDOS (PLDOS, defined in the next section) of scattering states impinging from the other side, with 
no requirement of spatial symmetry.  Thus, the PLDOS plays a more fundamental 
role than the LDOS.
Next, it turns out that there are significant deviations from the PLDOS that 
are highly sensitive to how far one is from a perfect transmission case.
Nevertheless, averaging over the tip position allows one to develop a 
quantitative method for extracting the PLDOS even in this regime.
Finally, it is shown that increasing the width of the tip reduces the 
quality of the PLDOS one can extract.

In Sec.\ \ref{sec:perturbative} the main results of the existing 
analytic perturbation theory\cite{jalabert2010,gorini2013} are summarized. 
The analytical derivation of the relationship between SGM and PLDOS for weak 
local tips is presented in Sec.\ \ref{sec:sgmldos_step} 
for the case of conductance steps and in Sec.\ \ref{sec:sgmldosunitary} 
for the case of perfect unit conductance. 
The corrections for non-perfect unit conductance are treated perturbatively in 
Sec.\ \ref{sec:sgmldos_closetounitary}.
A method for extracting the PLDOS and effectively disentangling first and 
second order contributions to the conductance response for imperfect transmission 
are given.
Numerical simulations of the second-order conductance correction dominant in the 
perfect transmission case are presented in Sec.\ \ref{sec:numericsg2_local} 
for the case of local tips and the full conductance correction is shown in 
Sec.\ \ref{sec:numericsg_local}.
The case of extended tips is discussed in Sec.\ \ref{sec:non-local}.
Some technical aspects related with the scattering states are relegated to 
Appendix \ref{sec:AppA}, and Appendix \ref{sec:AppB} establishes the link of 
a particular contribution to the SGM response with the LDOS.

\section{Partial local density of states in the scattering formalism}
\label{sec:pldos}

The spinless partial local density of states (PLDOS) for electrons impinging into the scatterer from lead $l$ can be defined by\cite{Buettiker1996,gramespacher1999}
\begin{equation}\label{ldossc}
\rho_{l\varepsilon}(\br)=2\pi \sum_{a=1}^{N}
|\Psi_{l,\varepsilon,a}(\br)|^2\, ,
\end{equation}
using the sub-ensemble of the basis of outgoing scattering states \eqref{allscats} incoming from lead $l$. $N$ is the number of propagating modes in the lead at the energy $\varepsilon$. 

The decomposition of the spinless local density of states (LDOS) $\rho_{\varepsilon}(\br)$ as
\begin{equation}
\label{pldos12}
\rho_{\varepsilon}(\br)= \rho_{1\varepsilon}(\br)+\rho_{2\varepsilon}(\br)\, ,
\end{equation}
valid for the two-lead case, naturally appears in scattering problems in which one is concerned with the response of the system to a small perturbation of the confining potential.\cite{gasparian1996} 
Such is the case of the SGM response, as well as that of the self-consistent treatment of electrical a.c.\ transport in mesoscopic systems.\cite{gramespacher1999}
The definition \eqref{ldossc} corresponds to an injectivity, \cite{gasparian1996,Buettiker1996,gramespacher1999} where the preselection of carriers is done by the incident lead $l$ from where they impinge into the scatterer.

Denoting by $M$ the number of open transmitting eigenchannels, the basis of scattering eigenfunctions \eqref{xis} gives the expressions for the PLDOS on the right and left, respectively, of the scatterer as 
\begin{subequations}
	\label{ldos} 
	\begin{align}
		\rho_{1\varepsilon}(\br)=2\pi \sum_{m=1}^{M}
			|\chi_{1,\varepsilon, m}(\br)|^2
		\, , \quad x >  \ 0 \ ,
		\label{ldos1}
		\\
		\rho_{2\varepsilon}(\br)=2\pi \sum_{m=1}^{M}
			|\chi_{2,\varepsilon, m}(\br)|^2
		\, , \quad x <  \ 0 \ .
		\label{ldos2}
	\end{align}
\end{subequations}
Since quite generally, $M \ll N$, and the transmitted parts of the scattering eigenstates \eqref{xis} are proportional to the diagonal elements of the transmission submatrices, the expressions \eqref{ldos} are considerably easier to evaluate than \eqref{ldossc}. However, it is important to keep in mind that the expressions \eqref{ldos1} and \eqref{ldos2} only describe the region opposite to the lead determining their PLDOS, and are not appropriate for obtaining the LDOS using Eq.\ \eqref{pldos12} since they refer to different regions of space. For instance, for a QPC embedded in a clean 2DEG,  $\rho_{1\varepsilon}(\br) \propto 1/|\br|$ far away from the QPC \cite{abbout2011}, while the LDOS is independent of $\br$.

\section{Perturbative results}
\label{sec:perturbative}

An analytical description of SGM in the presence of a strong tip is a challenging 
theoretical task. 
However, a perturbative approach\cite{jalabert2010} is tractable
in the weakly invasive case, where the tip-induced potential constitutes a small perturbation of the electrostatic potential seen by the electrons.
To begin, consider a weak tip potential 
$V_\mathrm{T}(\br)=v_\mathrm{T} f (\br -\br_\mathrm{T})$, where $f(\br)$ is 
a normalized function with $\int \dif\br  f(\br)=1$, which perturbs the system.
The change in the dimensionless (in units of $2 e^2 /h$) tip-position 
dependent conductance can be written as
\begin{equation}\label{taylor}
	g(\br_\mathrm{T}) = g^{(0)}+\delta  g(\br_\mathrm{T}) \, ,
\end{equation}
with
\begin{equation}\label{deltag}
\delta g (\br) = v_\mathrm{T} g^{(1)}(\br)+v_\mathrm{T}^2 g^{(2)}(\br) 
+O[v_\mathrm{T}^3]\, .
\end{equation}
The unperturbed conductance $g^{(0)}$ is given by the Landauer-B\"uttiker 
formula as the total transmission probability.  It can be expressed as a trace 
over the propagating modes 
\begin{equation}\label{eq:LB}
g^{(0)}=\mathrm{Tr}[t^\dagger t] = \sum_{m=1}^{M} {\cal T}_m^{2}
\end{equation}
in terms of the transmission submatrix $t$ of the unperturbed scattering matrix $S$ at the Fermi energy $\EF$ or the transmission eigenvalues ${\cal T}_m$ of the $M$ open eigenchannels (see Appendix \ref{sec:AppA}). 

The basis of the transmission eigenmodes is particularly suited to express the SGM conductance corrections. Assuming time-reversal invariance from here on, the two lowest-order 
corrections \cite{jalabert2010,gorini2013} are 
\begin{equation}	\label{fullg1_t_basis}
 	g^{(1)} =   \frac {4 \pi}{v_\mathrm{T}}
 	\mathrm{Tr}\left[\mathcal{R}\mathcal{T}
 	\mathrm{Im}\left\{\mathcal{U}^{21}\right\}\right] \, ,	
\end{equation}
\begin{widetext}
\begin{equation}\label{fullg2_t_basis}
 	\begin{aligned}
 	g^{(2)} = & -\frac {4 \pi^2}{v_\mathrm{T}^2}  \mathrm{Tr} \left[  
 	\mathcal{T}^2\ \mathcal{U}^{12}\mathcal{U}^{21}
 	- \mathcal{R}^2\ \mathcal{U}^{21}\mathcal{U}^{12} +
 	\mathcal{R}\mathcal{T} \mathrm{Re}\left\{ 
 	\mathcal{U}^{22}\mathcal{U}^{21}
 	-\mathcal{U}^{21}\mathcal{U}^{11} \right\} \right]
 	\\
 	&  -\frac {4 \pi}{v_\mathrm{T}^2}\sum_{\bl=1}^{2} 	
 	\mathcal{P} \int_{\varepsilon^\mathrm{t}_1}^{\infty} 
 	\frac{\dif\bepsilon}{\bepsilon-\EF} \
 	\mathrm{Tr}\left[ \mathcal{R}\mathcal{T} 
 	\mathrm{Im}\left\{\mathcal{U}^{2 \bl}(\EF,\bepsilon)
 	\mathcal{U}^{\bl 1}(\bepsilon,\EF)\right\}\right] \, . 
 	\end{aligned} 
\end{equation}
\end{widetext}
$\mathcal{R}$ and $\mathcal{T}$ are real diagonal reflection and 
transmission submatrices appearing in the polar 
decomposition \eqref{polarS} of $S$. The matrix elements  
\begin{equation}
\mathcal{U}^{\bar{l} l}_{\bar{m} ,m}(\bepsilon,\varepsilon)=
\int \chi^*_{\bl ,\bepsilon ,\bar{m}}(\br)V_\mathrm{T}(\br)
\chi^{\phantom{*}}_{l ,\varepsilon ,m}(\br) \dif \br
\end{equation} 
are those of the tip potential between two scattering eigenfunctions, 
where $l$, $m$ and $\varepsilon$ label the incoming lead, 
the channel number, and the electron energy of the scattering eigenfunction, 
respectively.
If the arguments of $\mathcal{U}^{\bar{l} l}$ are omitted, it is understood 
that both are taken at $\EF$, and the necessary matrix element tip position 
dependence on $\br_\mathrm{T}$ is assumed. 
The limiting integration energy $\varepsilon^\mathrm{t}_{1}$ is that of 
the lowest transverse energy and $\mathcal{P}$ stands for the principal part of the integral. 
The traces over the $N$ propagating modes in the leads in Eqs.\  \eqref{fullg1_t_basis} and \eqref{fullg2_t_basis} are dominated by  the contribution from the subspace of the $M$ open eigenmodes.

On a conductance plateau where the transmission is perfect,
$\mathcal{R}_m \mathcal{T}_m = 0$ for all eigenmodes $m$. 
There, the first order contribution \eqref{fullg1_t_basis} vanishes 
\cite{jalabert2010,gorini2013} 
and the SGM response is given by $v_\mathrm{T}^2g^{(2)}$; note that 
only one term above of $v_\mathrm{T}^2g^{(2)}$ survives as well. The relative importance of the linear and the quadratic SGM responses when moving between conductance plateaus and conductance steps of the QPC can also be affected by temperature, which mixes the two regimes, and can lead to an increase of the SGM response with increasing temperature as it was obtained in Ref.\ \onlinecite{abbout2011}.
    
Although it is not of direct experimental relevance, the case of a local 
tip $f(\br)=\delta (\br )$ is an interesting study case. In particular, 
the first-order conductance correction reduces to 
\begin{equation}\label{g1_ldos_0}
  g^{(1)} (\br_\mathrm{T}) =   4 \pi\sum_{m=1}^{M} 
  \mathcal{R}_m\mathcal{T}_m\mathrm{Im}\left\lbrace 
  \chi^*_{2,\EF,m}(\br_\mathrm{T}) 	\chi_{1,\EF,m}(\br_\mathrm{T})   
  \right\rbrace \, ,
\end{equation}
and the second-order correction for perfect transmission reduces to
\begin{equation}\label{corres}
    g^{(2)} (\br_\mathrm{T})  = - 4 \pi^2 \sum_{m,{\bar m}=1}^{M}
    |\chi_{2,\EF,{\bar m}}(\br_\mathrm{T})|^2 |\chi_{1,\EF,m}
    (\br_\mathrm{T})|^2 \, ,
\end{equation} 
where $M$ stands for the number of the partially open eigenchannels 
of the QPC in Eq.\ \eqref{g1_ldos_0} and perfectly open channels in 
Eq.\ \eqref{corres}. 
These expressions can be further simplified in cases exhibiting 
various kinds of symmetries and/or where the geometry allows for 
the evaluation of 
the scattering wave-functions.\cite{jalabert2010,gorini2013} 
   
\section{$g^{(1)}(\br_\mathrm{T})$ versus PLDOS in the conductance steps}
\label{sec:sgmldos_step}
  
Focusing first on a QPC setup without disorder, the asymptotic form 
of the scattering eigenfunctions can be used everywhere in the 2DEG, 
except in and very close to the constriction. 
The form \eqref{xis} enables expressing the product of scattering eigenfunctions impinging from different leads, in the 
first order correction \eqref{g1_ldos_0} due to a weak $\delta$-potential 
scanned in the right of the QPC, as
\begin{eqnarray}
  &&\chi^*_{2,\varepsilon,m}(\br)\chi^{\phantom{*}}_{1,\varepsilon,m}(\br)\nonumber \\
  &&\quad\quad =\mathcal{T}_m \left\{ \varrho_{2,\varepsilon,m}^{(+)\, 2} 
  (\br)+\mathcal{R}_m \varrho_{2,\varepsilon,m}^{(+)}(\br) 
  \varrho_{2,\varepsilon,m}^{(-)}(\br)\right\} \, .
\end{eqnarray}
Recalling  
$\varrho_{2,\varepsilon,m}^{(-)}(\br) = \varrho_{2,\varepsilon,m}^{(+)\, *}(\br)$, 
leads to 
\begin{equation}
  \mathrm{Im}\left\{ \chi_{2,\varepsilon,m}^*(\br)
  \chi^{\phantom{*}}_{1,\varepsilon,m}(\br) \right\}
  =\mathcal{T}_m 
  \mathrm{Im}\left\{\varrho_{2,\varepsilon,m}^{(+)\, 2}(\br)\right\} \, .
\end{equation}
From \eqref{scatst1} we have
$\varrho_{2,\varepsilon,m}^{(+)}(\br)=\chi_{1,\varepsilon,m}(\br)/\mathcal{T}_m$
for $x>0$ in the case of open modes ($\mathcal{T}_m \ne 0$). Thus, Eq.\ \eqref{g1_ldos_0} simplifies to
\begin{equation}\label{g1_ldos_00}
  g^{(1)} (\br_\mathrm{T})=   4 \pi\sum_{m=1}^{M} \mathcal{R}_m\mathrm{Im}\left\{ 
  \chi_{1,\EF,m}^2(\br_\mathrm{T}) \right\} \, .
\end{equation}
Denoting $\alpha_{l,\varepsilon,m}(\br)$ as the argument of 
$\chi_{l,\varepsilon,m}(\br)$, Eq.\ \eqref{g1_ldos_00} can be written
\begin{equation}\label{g1_ldos}
  g^{(1)} (\br_\mathrm{T})=   4 \pi
  \sum_{m=1}^{M} \mathcal{R}_m \sin[2 \alpha_{1,\EF,m}(\br_\mathrm{T})] 
  |\chi_{1,\EF,m}(\br_\mathrm{T})|^2	\, .   
\end{equation}
The sum over eigenmodes reduces to the contribution of the last one ($m=M$), 
which is the only partially open channel having $R_m>0$. 

In the case of a single open channel ($M=1$) there is a direct relation between the first-order conductance change and the PLDOS since, according to \eqref{ldos},
\begin{equation}
\label{g1_pldos}
  g^{(1)} (\br_\mathrm{T})= 
  2 \mathcal{R}_1 \sin[2 \alpha_{1,\EF,1}(\br_\mathrm{T})]
  \rho_{1\EF}(\br_\mathrm{T})	\, .   
\end{equation}
However, in the case of $M > 1$, the structure of the $m$-sum in Eq.~\eqref{g1_ldos} does not reduce to a simple relationship with $\rho_{1\EF}(\br)$.

In a disorder-free 2DEG, the prefactor $\sin(2\alpha_{1,\EF,1})$ of the SGM 
response \eqref{g1_pldos} is simply $\sin(2k_\mathrm{F}r+\alpha_0)$ with a constant phase $\alpha_0$, 
thus generating half Fermi wavelength, $\lambda_\mathrm{F}/2$, oscillations and a proportionality 
factor $2 \mathcal{R}_1$ between the spatial oscillation amplitude of the 
first order conductance correction in the first step and the PLDOS. 

In the case of a disordered structure, Eq.\ \eqref{g1_pldos} does not apply inside the disordered region, nevertheless if the disorder is weak and leads to small-angle forward scattering only, one can expect the structure of Eq.\ \eqref{g1_pldos} to mostly remain.  For example, the phase oscillation cannot have such a simple position-dependence strictly speaking, but a paraxial optical approximation~\cite{tomsovic2003} 
holds and a fairly regular radial phase behavior of nearly the same wavelength persists in the eigenfunctions.
In these circumstances, the explicit dependence of the SGM response on the phase of 
the scattering eigenfunction might be helpful in characterizing properties of 
the fluctuating potential in the 2DEG with further analysis. 

In general, the first-order conductance correction in tip-strength is not 
proportional to the PLDOS, even for the case of a $\delta$-tip.
In fact, $g^{(1)} (\br_\mathrm{T})$ is only local in the sense that 
$\mathrm{Im}\left\{ \chi_{1,\EF,m}^2(\br_\mathrm{T}) \right\}$ is 
the local information about the eigenfunction of the unperturbed system.  
However, in the case of a single partial mode the PLDOS provides an 
upper bound for the absolute value of the former, and the sinusoid term 
creates a fringing effect.  

For one-dimensional tight-binding systems the SGM response has been expressed 
in terms of the real part of the local Green function \cite{gasparian1996,pala2008} 
and thereby related to the LDOS.
We have checked that in the case of a one-dimensional chain the 
first-order conductance correction \eqref{g1_ldos_0} (and therefore also the 
relation \eqref{g1_pldos}) is consistent with the result of 
Refs.\ \onlinecite{gasparian1996,pala2008}. However, \eqref{g1_ldos_0} is more 
general and \eqref{g1_pldos} is expected to be valid whenever there is only 
one single partially open mode of the QPC, without being limited to strictly 
one-dimensional systems.

\section{Correspondence between $g^{(2)}(\br_\mathrm{T})$
and PLDOS for perfect transmission}
\label{sec:sgmldosunitary}

Symmetries have been shown to play a key role in the quest of 
identifying SGM maps with local properties.\cite{gorini2013} 
In particular, for a four-fold symmetric QPC operating in the regime of 
perfect transmission, the conductance 
change induced by a weak local tip in the absence of magnetic 
field has been shown to be proportional to the square of the LDOS, 
and also proportional to the local current density.
In the same framework, it has been pointed out \cite{szewcthesis} that the  
correspondence with the PLDOS holds even for asymmetric QPCs, provided that 
the conductance is set to the first plateau, as long as the system remains 
time reversal invariant. 

An important task, undertaken in this section, is the generalization of 
previous results to any conductance plateau of an arbitrary QPC
under the sole assumptions of time-reversal symmetry and a local tip.
To describe transport within the Landauer formalism,
the QPC can be treated as a scatterer centered at the origin $\br=0$. 
With the definitions of Appendix \ref{sec:AppA}, 
$\varphi^{(-)*}_{l,\varepsilon,m}(\br)=\varphi^{(+)}_{l,\varepsilon,m}(\br)$, 
and $\varrho^{(-)*}_{l,\varepsilon,m}(\br)=\varrho^{(+)}_{l,\varepsilon,m}(\br)$.  
Therefore, on the $m$-th conductance plateau, where ${\cal{R}}_m=0$,     
\begin{equation}\label{chi_TR}
\chi^{\phantom{*}}_{2,\varepsilon,m}(\br)=\chi^*_{1,\varepsilon,m}(\br)
\end{equation} 
in the 2DEG on both sides of the QPC.
Using this relationship in the second order correction \eqref{corres} leads to
\begin{equation}
\label{corres2}
g^{(2)}(\br_\mathrm{T}) = - \rho^2_{1\EF}(\br_\mathrm{T}) \, ,
\end{equation}
for $\br_\mathrm{T}$ at the right of the QPC.

Unlike the relation for the first step, which is linear in the PLDOS and fringed in space, perfect transmission on any plateau leads to a quadratic dependence on the PLDOS without fringing. Interestingly, no spatial symmetry is required for the correspondence \eqref{corres2} in the considered regime of conductance quantization. Nevertheless, a perfect conductance quantization with exact unit transmission is a regime difficult to reach in experiments with real QPCs.  

\section{$g^{(2)}(\br_\mathrm{T})$ versus PLDOS near perfect transmission}
\label{sec:sgmldos_closetounitary}

In Sec.\ \ref{sec:sgmldosunitary} perfect transmission is assumed 
in order to establish the correspondence between the second order 
conductance correction and the PLDOS. 
Here that condition is relaxed. 
Beyond the unity case of perfect conductance quantization where all $\mathcal{R}_m=0$, the first-order 
correction \eqref{g1_ldos} is nonzero, and all terms of the second-order 
correction $g^{(2)}$ in Eq.\ \eqref{fullg2_t_basis} must be considered.

Begin with the situation of transmission slightly below the unity case
on the $M$\textsuperscript{th} conductance plateau, 
where the transmission of the highest open channel $M$ is not perfect.
The expressions of the scattering eigenstates 
\eqref{scatst1} and \eqref{scatst2} can be used to find that
\begin{equation}\label{chi2_vs_chi1}
 \chi^{\phantom{*}}_{2,\varepsilon,m}(\br)=\frac{1}{\mathcal{T}_m}\left(1+\mathcal{R}_m 
 e^{2 i\alpha_{1,\varepsilon,m}(\br)}\right)\chi^*_{1,\varepsilon,m}(\br)
\end{equation} 
for an open mode at the right of a generic QPC.
By inserting \eqref{chi2_vs_chi1} into Eq.\ \eqref{fullg2_t_basis} 
(where the last term is related to a Hilbert transform of the 
density of states [see Appendix \ref{sec:AppB}]), and only
keeping the lowest order terms in $\mathcal{R}_m$, $g^{(2)}$ reads
\begin{widetext}
\begin{equation}\label{r_perturbation1}
 g^{(2)}(\br_\mathrm{T}) =  
    - 2\pi \rho_{1 \EF}\sum_{m=1}^{M}|\chi_{1,\EF,m} (\br_\mathrm{T})|^2
    \left(1+2\mathcal{R}_m
    \left\{ 
 	\cos\left[2\alpha_{1,\EF,m}(\br_\mathrm{T})\right] 
 	+\eta_{\EF}(\br_\mathrm{T})
 	\sin\left[2\alpha_{1,\EF,m}(\br_\mathrm{T})\right] \right\}\right)	
 	\, , 
\end{equation}
where  
\begin{equation}\label{eq:eta}
 \eta_{\EF}(\br)=\frac{1}{\pi}
 {\cal P}\int_{\varepsilon^\mathrm{t}_1}^{\infty} 
 \frac{\dif\bepsilon}{\bepsilon-\EF}\frac{\rho_{\bepsilon}(\br)}{2\rho_{1 \EF}(\br)}
\end{equation}
for positions $\br$ to the right of the QPC. Notice that the relation of 
the LDOS to the imaginary part of the diagonal Green function 
$\mathcal{G}_\varepsilon(\br,\br)$ implies 
$\eta_{\EF}(\br)=
-\mathrm{Re}\mathcal{G}_{\EF}(\br,\br)/(2\pi\rho_{1 \EF}(\br))$. 
Taking $\mathcal{R}_m=0$ for all $m < M$ gives
\begin{equation}\label{r_perturbation2}
 g^{(2)}(\br_\mathrm{T}) = - \rho_{1 \EF}^2
    - 4\pi \mathcal{R}_M \rho_{1 \EF}|\chi_{1,\EF,M} (\br_\mathrm{T})|^2
    \left\{\cos\left[2\alpha_{1,\EF,M}(\br_\mathrm{T})\right] 
 	+\eta_{\EF}(\br_\mathrm{T})
 	\sin\left[2\alpha_{1,\EF,M}(\br_\mathrm{T})\right] \right\}	\, , 
\end{equation}
and the small reflection amplitude is linked to the deviation from 
unit conductance by $\Delta g = \mathcal{R}_M^2$, where 
$\Delta g=M - g^{(0)}$ quantifies the departure from unit transmission 
on the $M$\textsuperscript{th} plateau. 
In the case of unit transmission one has $\mathcal{R}_M=0$, 
and \eqref{r_perturbation2} reduces to \eqref{corres2}.  
For completeness, in the same regime Eq.\ \eqref{g1_pldos} 
can be rewritten as
\begin{equation}
\label{g1Mterm}
g^{(1)}(\br_\mathrm{T}) = 4 \pi \mathcal{R}_M |\chi_{1,\EF,M}(\br_\mathrm{T})|^2 \sin[2 \alpha_{1,\varepsilon,M}(\br_\mathrm{T})]  \ ,
\end{equation}
which has similarities in its form with respect to the correction terms 
for $g^{(2)}(\br_\mathrm{T})$.  Recall however, the corresponding 
conductance correction varies linearly with the strength of the tip 
potential unlike for $g^{(2)}(\br_\mathrm{T})$.

In the case of transmission just above the unity case with low transmission 
$\mathcal{T}_{M+1}$ through the QPC 
mode $M+1$, a similar procedure, 
assuming $\mathcal{R}_m=0$ for all $m \le M$ and keeping only the lowest 
terms in $\mathcal{T}_{M+1}$ yields
\begin{eqnarray}\label{r_perturbation3}
 g^{(2)}(\br_\mathrm{T}) &=& - \rho_{1 \EF}^2 
 + 2\pi \mathcal{T}_{M+1}^2 \left|\varrho_{2,\EF,M+1}^{(-)}(\br_\mathrm{T})\right|^2 
 \times
  \nonumber \\
    &\times & 
    \left\{\rho_{1 \EF} 
    + 4\pi \left| \varrho_{2,\EF,M+1}^{(-)} (\br_\mathrm{T})\right|^2
    \left( 1+\cos\left[2\alpha_{1,\EF,M+1}(\br_\mathrm{T})\right]\right)^2
    -2\rho_{1 \EF} \eta_{\EF}(\br_\mathrm{T})
    \sin\left[2\alpha_{1,\EF,M+1}(\br_\mathrm{T})\right]
 	 \right\}\, .
\end{eqnarray}
\end{widetext}
The small transmission in the QPC channel $M+1$ causes departures 
from \eqref{corres2} that are expected to be proportional to 
$\mathcal{T}_{M+1}^2$.

However, in a real system slightly above integer dimensionless 
conductance, the small transmission of the $M+1$\textsuperscript{st} 
channel can coexist with an imperfect transmission of the 
$M$\textsuperscript{th} channel, 
$\Delta g= \mathcal{R}_M^2 - \mathcal{T}_{M+1}^2 $, and the 
departure from \eqref{corres2} has contributions from both channels, which are difficult to separate in the numerical work.
To avoid this complication, we concentrate in the following on the case of 
positive $\Delta g$, at positions on the conductance plateau where the 
opening of the next channel is exponentially suppressed and thus 
negligible. 

It is worth emphasizing a few features of the expressions contained in 
Eqs.\ (\ref{r_perturbation2}, \ref{g1Mterm}).  The scale of the deviations from the square of the PLDOS is greatly magnified by being proportional to the square root of $\Delta g$ as opposed to being linear.  In other words the approach to the perfect transmission case is rather slow with respect to the limit $\Delta g \rightarrow 0$, and even tiny imperfections produce highly visible deviations.  Nevertheless, all the deviations oscillate about zero with a wavelength on the order of $\lambda_F/2$, and a spatial averaging over a region $\lambda_F/2 \times \lambda_F/2$ results in a near uniform distribution of angles $\alpha$ over $2\pi$, giving a means for the near elimination of the correction terms in \eqref{r_perturbation2}.
Thus, though with reduced spatial resolution, it is still possible to cleanly extract the PLDOS.  The PLDOS is not proportional to the LDOS in this case, and the distinction matters.

Furthermore, since the contribution of $g^{(1)}(\br_\mathrm{T})$ to $\delta  g(\br_\mathrm{T})$ is linearly proportional to the tip strength $v_\mathrm{T}$ and the contribution of $g^{(2)}(\br_\mathrm{T})$ quadratic, measurements with two well chosen values of $v_\mathrm{T}$ would be sufficient to separate out the contributions from 
Eqs.\ (\ref{r_perturbation2}) and (\ref{g1Mterm}); with a few more tip strength measurements per tip site, noise and other inaccuracies could be overcome in the separation as well.  In the event that $ |\chi_{1,\EF,M}(\br_\mathrm{T})|^2$ mostly varies slowly on the scale of $\lambda_F$, then probability densities due to individual eigenstates and the spatial behavior of $\alpha$ could be extracted as well. Given that $\eta(\br)$ is related to the phase of the real part of the diagonal Green function, in an ideal situation, it could be extracted also.


In order to quantify the departures of $g^{(2)}(\br_\mathrm{T})$ 
from the perfect case, introduce the 
ratio between the coefficient of the second order SGM correction and 
the square of the PLDOS
\begin{equation}
 \kappa(\br_\mathrm{T})
 =-\frac{g^{(2)}(\br_\mathrm{T})}{\rho_{1 \EF}^2(\br_\mathrm{T})} \, .
\end{equation} 
If the unperturbed conductance $g^{(0)}$ is just below that of $M=1$, 
and the sum over QPC eigenmodes is restricted to $m=1$, then
\begin{equation}\label{kappa}
 \kappa(\br)=1+2\sqrt{\Delta g}\left\{\cos\left[2\alpha(\br)\right] 
 +\eta(\br)\sin\left[2\alpha(\br)\right]\right\}\, .
\end{equation}
The indices of $\alpha$ and $\eta$ are omitted; it is understood 
that $\alpha=\alpha_{1,\EF,1}$ and $\eta=\eta_{\EF}$. 

As mentioned above, even fairly local spatial averaging approximately yields 
$\overline{\kappa}=\langle\kappa(\br)\rangle=1$. Interest is therefore in the quantity $\kappa-1$.
Similar to the case of the first-order SGM correction at a conductance step, 
discussed in Sec.\ \ref{sec:sgmldos_step}, the above relationship provides 
bounds for the possible values of the ratio $\kappa$, 
\begin{equation}\label{kappa_limits}
| \kappa-1| \le  2\sqrt{\Delta g}\sqrt{1+\eta_\mathrm{max}^2} \, ,
\end{equation}
where $\eta_\mathrm{max}$ is the maximum value of $|\eta(\br)|$. \textit{A priori},
$\eta_\mathrm{max}$ is not known, but if not extracted as described, it can be obtained by direct numerical computation of the scattering wavefunctions (see Sec.\ \ref{sec:numericsg2_local})
or estimated from simple setups, like that of an abrupt QPC, where
the analytical form of the scattering wave-functions is known. \cite{gorini2013} 
The maximum value of $\eta$ occurs in regions where the PLDOS is weak, and can in general approach infinity.  It's actual value depends on the problem and region under consideration.  In one numerical example given ahead, its maximum is of the order of $60$.

Another interesting quantity is the variance of $\kappa -1$ given by
\begin{equation}\label{sigma}
	\sigma^2 = 2\Delta g(1+\overline{\eta^2}) \, ,
\end{equation}  
where $\overline{\eta^2}$ is the average value of $\eta^2$ in the scan region.

\begin{figure}
\includegraphics[width=\linewidth]{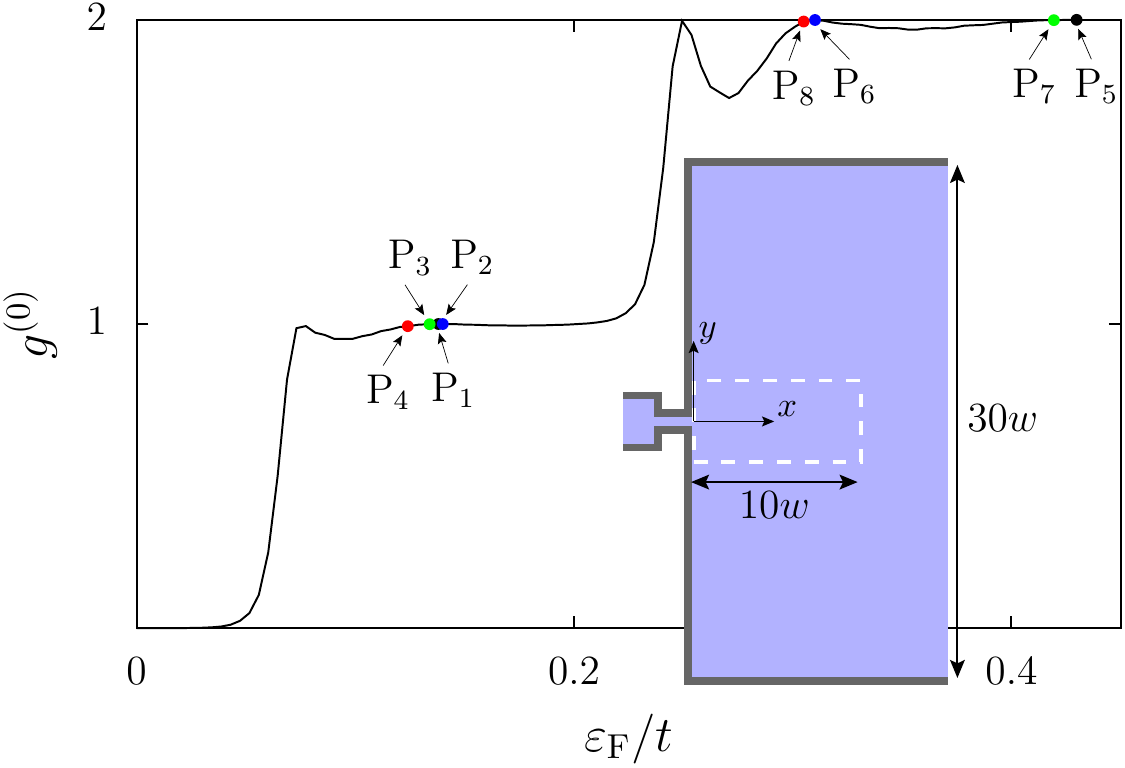}
\caption{\label{figure1} 
The conductance of the QPC defined in a tight binding lattice 
with lattice parameter $a$ and hopping $t$ as a function of 
Fermi energy. 
The inset shows the geometry of the QPC. The width and length of the 
narrow channel are $w=11a$ and $L=19a$, respectively. 
The points $\mathrm{P}_1$--$\mathrm{P}_8$ indicate the Fermi 
energies and unperturbed conductances 
at which the statistics of Sec.\ \ref{sec:numericsg2_local}
have been performed using tip positions inside the dashed white 
rectangle.}
\end{figure}

\section{$g^{(2)}(\br_\mathrm{T})$ versus PLDOS for local tips : simulations}
\label{sec:numericsg2_local}

In order to test our analytical approach and go beyond the above described 
perturbation theory we performed numerical simulations using the quantum 
transport package \textsc{kwant} \cite{kwant-paper} that is based on the recursive 
Green function method. \cite{lee1981}  It can be used to calculate $\delta g (\br)$ as a direct 
subtraction, and $g^{(1)}(\br_\mathrm{T})$ or $g^{(2)}(\br_\mathrm{T})$ by constructing numerical derivatives with respect to $v_\mathrm{T}$. 
 
\begin{figure*}
\includegraphics[width=0.9\linewidth]{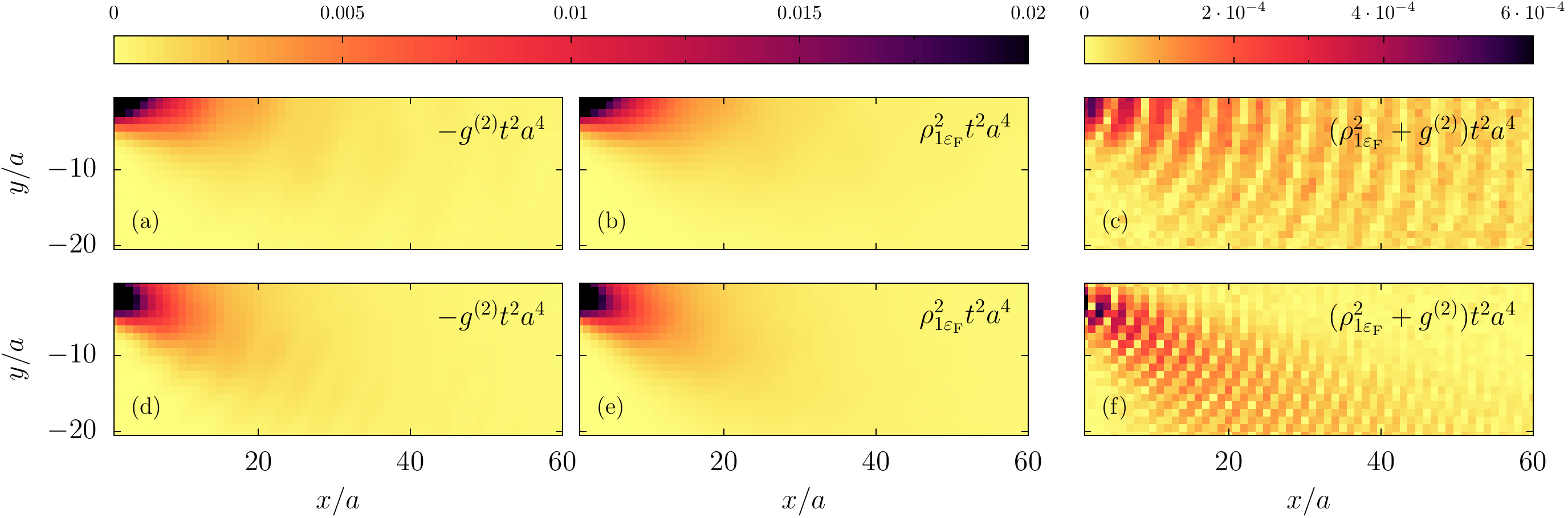}
\caption{\label{figure2}
Left column: $-g^{(2)}$ (with the energy and length units introduced 
through the hopping integral $t$ and the spatial tip extension $a^2$) \textit{vs}.\ 
the tip position for the first (a) and second (d) plateaus 
(points $\mathrm{P}_1$ and $\mathrm{P}_5$ in Fig.\ \ref{figure1}, respectively) 
Central column: the square of the 
PLDOS for the same points on the first (b) and second (e) plateau. 
Right column: difference between the two first columns. 
The QPC is situated at the upper left corner of the figures.}
\end{figure*}
In our simulations the 2DEG is discretized on a tight binding network with 
lattice parameter $a$ and a hopping integral $t=\hbar^2/(2{m^*}a^2)$, 
$m^*$ being the electron's effective mass. We chose an abrupt constriction 
defined by a hard-walled 
square well of width $w=11 a$ and length $L=19 a$ attached to two 
semi-infinite leads, sketched in the inset of Fig.\ \ref{figure1}. 
In order to optimize the computational time the left lead is narrowed. 
Figure \ref{figure1} shows the dimensionless conductance through the QPC as 
a function of the Fermi energy of the incoming electrons. 
As the latter is increased the QPC's conductance increases in steps 
of unit height. The structures on the plateaus are due to the abruptness of 
the QPC that lead to Fabry-Perot like oscillations within 
the constriction. \cite{szafer1989}

\subsection{Local correspondence for perfect transmission}

In order to address this regime, consider the analytically predicted 
relationship \eqref{corres2} between the second-order conductance 
correction $g^{(2)}(\br_\mathrm{T})$ for a $\delta$-tip and the 
PLDOS for perfect conductance. On the tight binding lattice, the 
$\delta$-tip is modeled as an additional 
on-site energy $\varepsilon_\mathrm{T}$ on a single site, corresponding 
to a tip area of $a^2$ and thus $v_\mathrm{T}=\varepsilon_\mathrm{T}a^2$.  This strength is varied so as to extract $g^{(2)}(\br_\mathrm{T})$.  The Fermi energies are chosen on the first and second plateaus for which the values of the unperturbed conductances $g^{(0)}$ are very close to perfect transmission
with $|\Delta g| < 10^{-5}$ (points $\mathrm{P}_1$ and $\mathrm{P}_5$ in Fig.\ \ref{figure1}).  The corresponding Fermi wavelengths are $\lambda_\mathrm{F}=16.8a$ and $\lambda_\mathrm{F}=9.4a$, respectively.  
The resultant conductance responses are shown in Fig.\ \ref{figure2}, where $g^{(2)}(\br_\mathrm{T})$ is compared to $-\rho_{1 \EF}^2$ for the first plateau case in panels (a) and (b) and likewise for the second plateau case in panels (d) and (e).  The correspondence is excellent as expected given the regime of the calculation.  This is illustrated in panels (c) and (f), which show the differences,$\left[\rho_{1 \EF}^2 +  g^{(2)}(\br_\mathrm{T})\right]$, respectively, for the two plateaus.  The differences are quite small as is expected and they show the $\lambda_\mathrm{F}/2$ 
oscillations, which are characteristic of the correction terms for imperfect transmission. 

\subsection{Departures from local correspondence for imperfect transmission}
\label{sec:numerics_closetounitary}

\begin{figure}
\includegraphics[width=\linewidth]{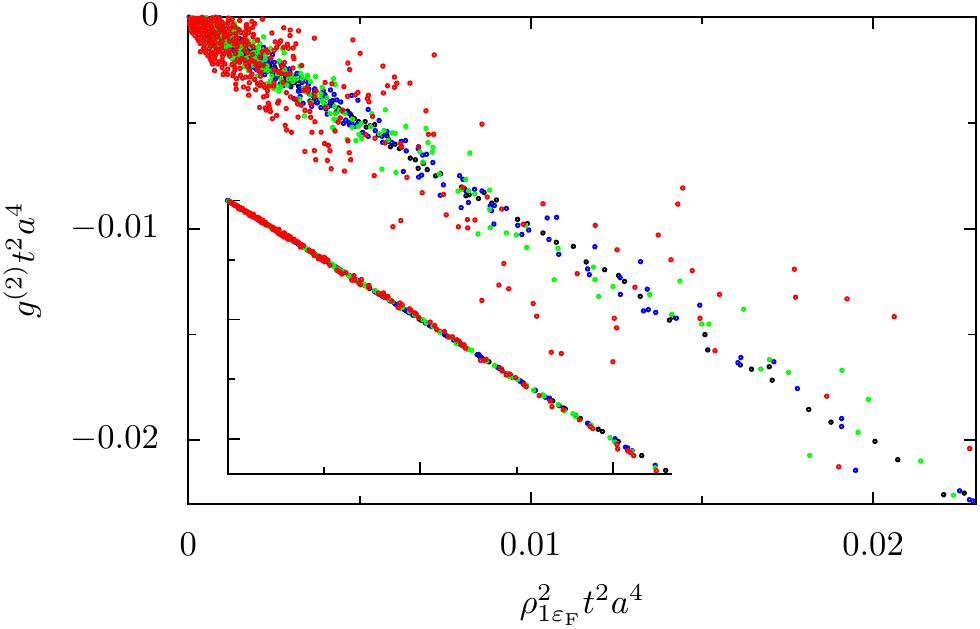}
\caption{\label{figure3}
Second order SGM correction \textit{vs}.\ $\rho_{1\EF}^2$
at random sampled tip positions in the scanned region for different values 
of the unperturbed conductance on the second plateau (points $\mathrm{P}_5$,  
$\mathrm{P}_6$, $\mathrm{P}_7$, and $\mathrm{P}_8$ in Fig.\ \ref{figure1}).
The corresponding departures from the quantized value are 
$\Delta g = 8\times 10^{-6}$, $5\times 10^{-4}$, $10^{-3}$, and $6\times 10^{-3}$ 
for the black, blue, green and red points, respectively. 
Inset: the same data are presented after a spatial average 
over a disk of radius of $\lambda_\mathrm{F}/2$, exhibiting a clear data 
collapse.}	
\end{figure}

It is shown in Sec.\ \ref{sec:sgmldos_closetounitary} that the precise local 
correspondence between the second-order SGM correction and the PLDOS squared degrades away 
from perfect transmission. We now present a quantitative numerical analysis of the 
departure from local correspondence for the example of the second conductance 
plateau of the QPC. Similar results can be obtained on other plateaus.
Figure \ref{figure3} presents the values of $g^{(2)}(\br_\mathrm{T})$ and $\rho_{1\EF}^2$
at different points of the scanned region inside the white dashed rectangle 
shown in the inset of Fig.\ \ref{figure1}.  The region of length $10w$ has been 
chosen so as to contain points close to the QPC and at larger distances.  This region width is small as compared to the width of the 2DEG ($30w$), and additional lateral leads on the full length at the right of the QPC are used
in order to avoid finite size effects. 

\begin{figure}
\includegraphics[width=\linewidth]{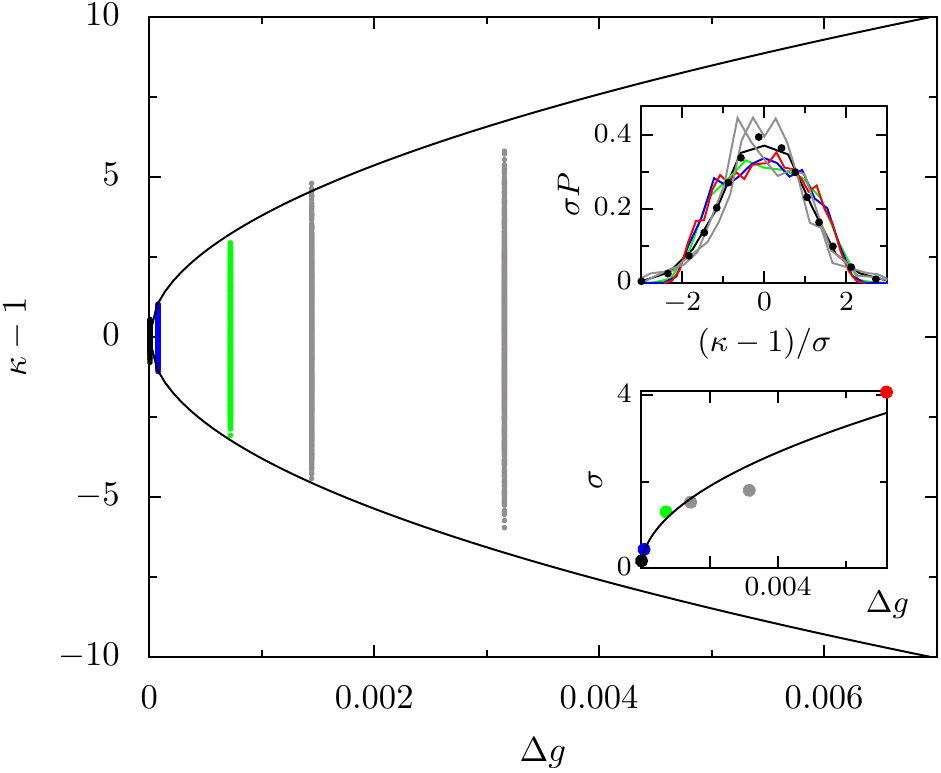}
\caption{\label{figure4}
$\kappa -1$ is plotted \textit{vs}.\ the departure from 
perfect transmission $\Delta g$, when a wide region in the right 
side of the QPC is sampled. The results for the clean structure 
of Figs.\ \ref{figure2} and \ref{figure3}, 
$\Delta g = 6\times 10^{-6}$ 
(black; $\mathrm{P}_1$ in Fig.\ \ref{figure1}), 
$\Delta g = 8\times 10^{-5}$ (blue; $\mathrm{P}_2$), 
$7\times 10^{-4}$ (green; $\mathrm{P}_3$) are presented, but those for 
$\Delta g = 7\times 10^{-3}$ (red; $\mathrm{P}_4$) are out of the scale 
of the main figure. 
The data corresponding to two different disorder configurations are 
represented by the gray distributions. 
The  black solid lines show the 
analytical bounds $\kappa_\pm$ of Eq.\ \eqref{kappa_limits} taking 
$\eta_\mathrm{max} = 60$. 
Upper inset: the probability density of $\kappa -1$. 
The color code is the same as in the main figure. For comparison, 
the dotted line shows a Gaussian probability density. 
Lower inset: the corresponding standard deviation \textit{vs}.\ $\Delta g$. 
The black solid line corresponds to the analytical expression \eqref{sigma} 
of $\sigma$ with $\overline{\eta^2}=(\eta_\mathrm{max}/2)^2$.}
\end{figure} 
The data shown in Fig.\ \ref{figure3} confirm that the exact 
point-by-point local correspondence is progressively broken as 
$|\Delta g|$ increases. Close to the perfect transmission condition, 
for the case with $\Delta g = 8\times 10^{-6}$ ($\mathrm{P}_5$ in 
Fig.\ \ref{figure1} with scans depicted in the lower panels of 
Fig.\ \ref{figure2}), the equivalence between 
$-g^{(2)}(\br_\mathrm{T})$ and the square of the PLDOS is attained 
(black dots).
For other points of the unperturbed conductance shown in 
Fig.\ \ref{figure1}, 
$\mathrm{P}_6$ with $\Delta g = 5\times 10^{-4}$ (blue), 
$\mathrm{P}_7$ with $10^{-3}$ (green) and $\mathrm{P}_8$ with 
$6 \times 10^{-3}$ (red), 
the sampled points exhibit progressively wider distributions around 
the equivalence \eqref{corres2}. The distributions are displayed in 
Fig.\ \ref{figure4}, where $\kappa-1$ is plotted for different Fermi 
energies on the first plateau 
($\mathrm{P}_1$, $\mathrm{P}_2$, $\mathrm{P}_3$, and $\mathrm{P}_4$ 
in Fig.\ \ref{figure1}), labeled by the value of $\Delta g$. 
In agreement with our analytical findings of the previous section, 
the average value of $\kappa$ remains equal to one, but the width of the 
distribution drastically increases with $\Delta g$ within the bounds 
$\kappa_\pm$ established in Eq.\ \eqref{kappa_limits} (solid lines) 
using the value $\eta_\mathrm{max}=60$ of the abrupt QPC. 

The probability density of $(\kappa-1)/\sigma$ is shown in the upper 
inset of Fig.\ \ref{figure4}, for the same positions on the first 
conductance plateau. 
The rescaling by the variance collapses the probability densities for 
all the values of $\Delta g$ to approximately a universal Gaussian 
form (dotted line).  
The analytical result of \eqref{sigma} for the standard deviation 
$\sigma$ of the ratio $\kappa$ from its mean value ($\kappa =1$), 
is evaluated using the assumption
$\overline{\eta^2}=(\eta_\mathrm{max}/2)^2$, and is shown to agree 
with the numerical results (lower inset of Fig.\ \ref{figure4}). 

\begin{figure*}
\includegraphics[width=\fwidth]{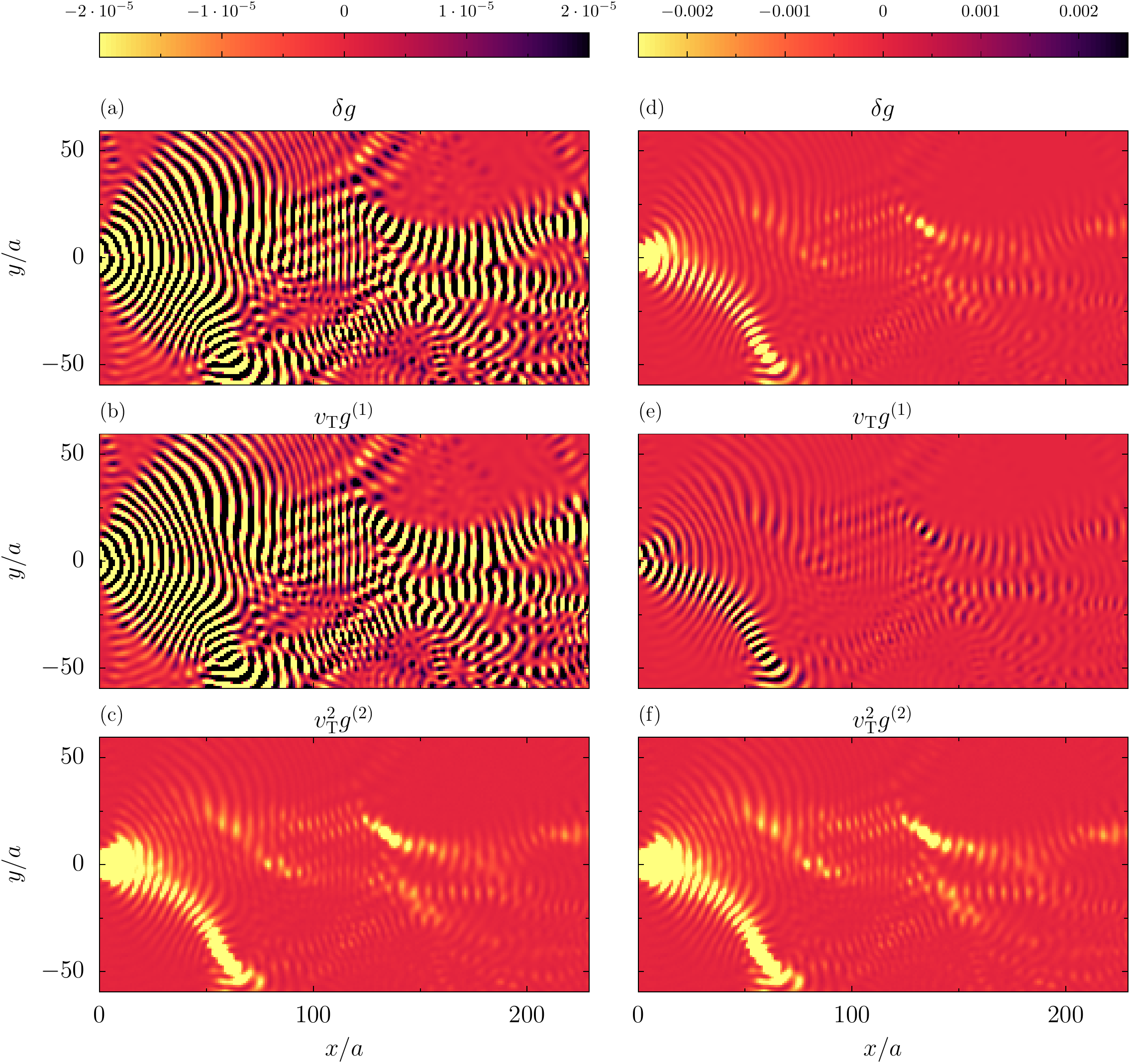}
\caption{\label{figure5}
SGM response for two tip strengths, $v_\mathrm{T}=\EF a^2/4$ 
(left column) and $v_\mathrm{T}=3\EF a^2$ (right column) with 
$\Delta g=1.3\times 10^{-4}$ on the second conductance plateau of 
a QPC in a disordered 2DEG.  
Ordered vertically for each case, the quantities plotted are: 
full response $\delta g(\br_\mathrm{T})$, first correction 
$g^{(1)}(\br_\mathrm{T})$, and second correction 
$g^{(2)}(\br_\mathrm{T})$.  The changing nature and relative balance 
of the different order terms is clearly visible. The weaker tip strength 
is expected to be dominated by the first order term in the left column, 
but not so for the stronger tip strength in the right column.}
\end{figure*}
The possible connection of SGM response with local properties needs 
to be extended to the realistic situation where the QPC is surrounded 
by a disordered 2DEG.
Though it is difficult to treat this case analytically because 
the asymptotic form of the scattering wave-functions is attained 
only beyond the region of disorder far from the QPC, the 
incorporation of disorder in the numerically tackled model is 
straightforward.
We assume the disorder to be due to randomly distributed donor atoms in 
a plane situated at a distance $z=10a$, with a concentration of 
$N_d=4\times 10^{-4} a^{-2}$. By taking $a=\unit[5]{nm}$, $N_d$ is equal 
to $\unit[10^{12}]{{cm}^{-2}}$, which is a realistic value for a high 
mobility 2DEG, and corresponds to elastic and transport mean free paths 
of $1$ and $\unit[52]{\mu m}$, respectively. 
The two vertical gray lines in Fig.\ \ref{figure4} correspond to 
samplings of different disorder configurations, resulting in small 
departures from unit transmission, which are quantified by the values 
of $\Delta g$. 
Thus, disordered QPCs, as well as clean ones, have departures from the 
local relation between $-g^{(2)}(\br_\mathrm{T})$ and the PLDOS squared 
that are uniquely governed by the crucial parameter $\Delta g$.

\subsection{Locally averaged correspondence for local tips}

\begin{figure*}
\includegraphics[width=\fwidth]{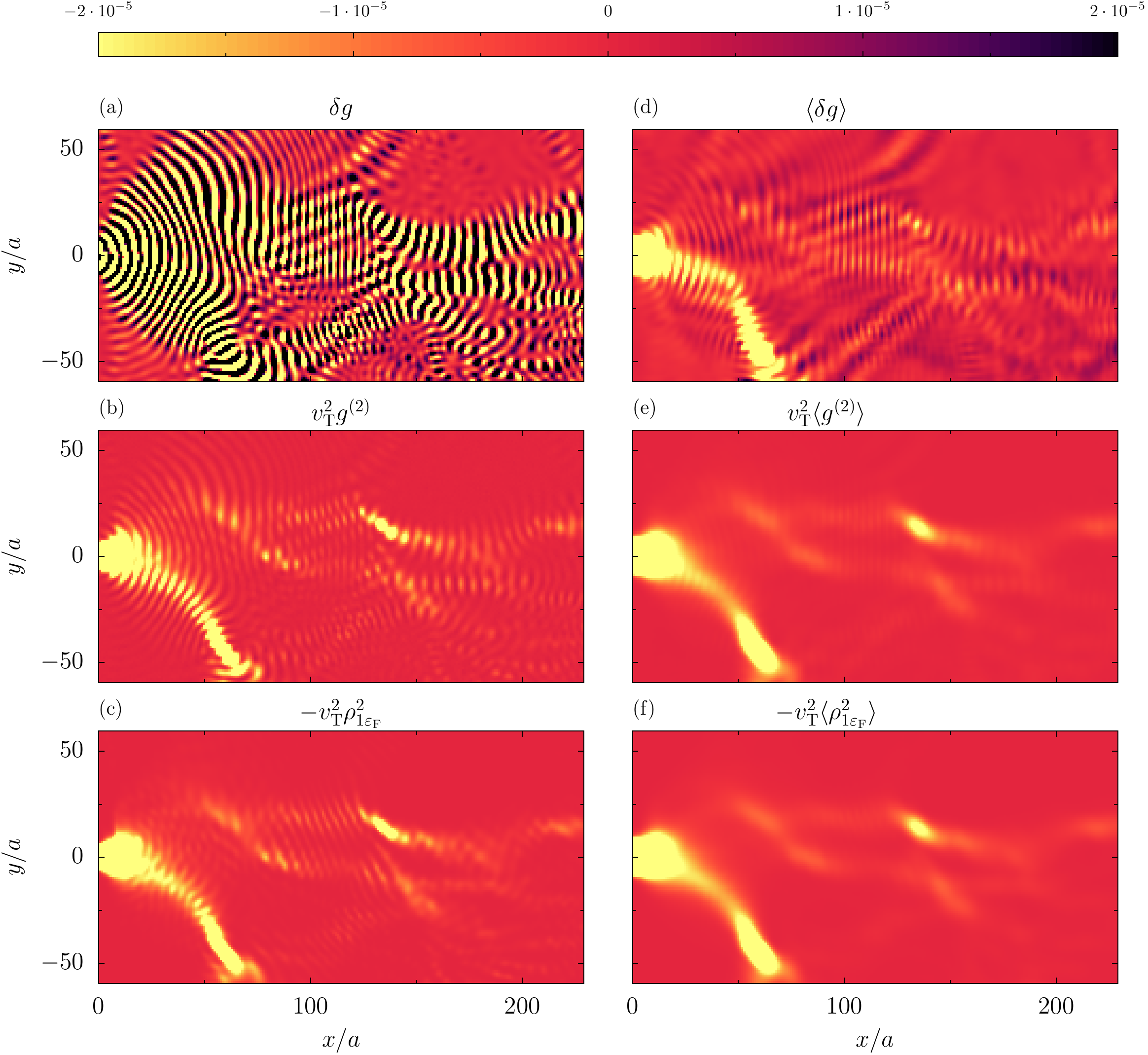}
\caption{\label{figure6}
Extracting an accurate PLDOS squared from the full SGM response 
in the weakly invasive regime for the disorder configuration of 
Fig.\ \ref{figure5}, for the weaker tip strength 
$v_\mathrm{T}=\EF a^2/4$:
(a) $\delta g(\br_\mathrm{T})$ on the right side of the QPC ; 
(b) the quadratic tip dependence portion of $\delta g(\br_\mathrm{T})$; 
(c) the negative of the squared PLDOS, $-\rho_{1 \EF}^2$. 
In (d), (e), and (f), respectively, the data of panels (a), (b), and (c) 
have been averaged over a disk of diameter $\lambda_\mathrm{F}/2$.}
\end{figure*}
Sections \ref{sec:sgmldos_closetounitary} and 
\ref{sec:numerics_closetounitary} show that even small deviations from 
perfect conductance drastically alter the SGM-PLDOS correspondence.
However, according to Eq.\ \eqref{kappa} and the calculations of the 
(Fig.\ \ref{figure3}) inset, the average of $\kappa$ is equal to unity.
The precise $\kappa$ values though should fluctuate in a quasi-random 
way with a standard deviation scaling as the square root of $\Delta g$.   
Such a behavior is the signature of the $\lambda_\mathrm{F}/2$-wavelength 
oscillations in the SGM response occurring in the clean case, which is 
modified in the presence of disorder.
Nevertheless, as discussed in Sect.\ \ref{sec:sgmldos_closetounitary},
the oscillations should self cancel once averaged over a domain of 
length scale as short as $\lambda_\mathrm{F}/2$ in both directions 
of the plane.  In order to verify this interpretation, the numerically 
obtained values are averaged over a disk of radius of 
$\lambda_\mathrm{F}/2$. As illustrated in the inset of 
Fig.\ \ref{figure3}, the averaging results in a data collapse 
yielding the equivalence between 
$\langle -g^{(2)}(\br_\mathrm{T})\rangle$ and 
$\langle \rho_{1\EF}^2(\br_\mathrm{T})\rangle$, 
even in the case of imperfect unit transmission. 
The recovery of the SGM-PLDOS correspondence upon averaging shows 
that there is a global structural correspondence with a 
characteristic length scale given by the Fermi wavelength.  
However, this correspondence is found for a local tip and
only between the PLDOS squared and the second order correction.

A finite temperature also has a tendency to reduce the fringes with 
period $\lambda_\mathrm{F}/2$ that are the main deviations from the 
SGM-PLDOS correspondence. Though the related mechanism is an energy 
average, very different from the spatial average proposed above, 
it might still be possible that a moderate temperature helps to 
improve the extraction of the PLDOS from SGM data.   

\section{Full SGM response for local tips}
\label{sec:numericsg_local}

\textit{A priori}, from an experimental point of view, the relationship between 
the various order terms and the full conductance change is not obvious.  
Even for weakly imperfect transmission somewhere on a plateau, 
depending on the tip strength, the full SGM response may depend not 
just on the leading second order term, but also crucially on the first 
and the other higher order terms.
Thus, $\delta g(\br_\mathrm{T})$ can vary considerably as a function 
of the tip strength for less than perfect transmission cases, which 
would most often be the case in experiments.
This is illustrated in Fig.\ \ref{figure5}, where 
$\delta g(\br_\mathrm{T}),g^{(1)}(\br_\mathrm{T}),g^{(2)}(\br_\mathrm{T})$ 
are plotted for two different tip strengths. 
The longer system treated here, in comparison with the simulations of 
Fig.\ \ref{figure2}, is numerically more demanding and thus the 
width of the 2DEG at the right of the QPC is limited to $20w$.  
The specific example illustrated is on the second plateau of the 
quantized conductance where $\Delta g = 1.3\times 10^{-4}$ using 
tip strengths of $v_\mathrm{T}=\EF a^2/4$ and $v_\mathrm{T}=3\EF a^2$.
The characteristic branching behavior of the fringes due to 
disorder\cite{topinka2001} is observed.
The changing nature of the full SGM response and its relationships 
with the linear and quadratic parts of the response are clearly seen.

Continuing to restrict ourselves to the weakly invasive regime, if the 
goal were to extract a local quantity, in this case, the square of the 
PLDOS, two operations would greatly enhance the quality of the analysis.  
The first is to make a few measurements with different tip strengths.  
Depending on the accuracy of the measurements or ambient noise, 
this would allow one to separate linear, quadratic, or even higher 
order variations with respect to tip strength.
The quadratic dependent response is the one related to the PLDOS 
squared; see Eq.\ \eqref{r_perturbation2}.
Second, one would average the data over a region of sidelength 
or radius $\lambda_\mathrm{F}/2$. 
Consider the weak tip strength case illustrated in Fig.\ \ref{figure5}.  
There, the first order term dominates the full SGM response 
$\delta g(\br_\mathrm{T})$. Nevertheless, extracting first the 
quadratic tip dependent part of the full response before averaging 
leads to a much more accurate extraction of the PLDOS squared. 
This is illustrated in Fig.\ \ref{figure6}.  
In the first row, $\delta g(\br_\mathrm{T})$ is shown with its 
locally averaged image to the right. In the next row, the quadratic 
tip dependence is deduced first, and then averaged.
Finally in the bottom row, the negative of the squared PLDOS is 
plotted along with its average. 
The improvement in the correspondence of the quadratic portion of 
$\delta g(\br_\mathrm{T})$ relative to the full response to the 
average PLDOS is quite striking.

\begin{figure}
\includegraphics[width=\linewidth]{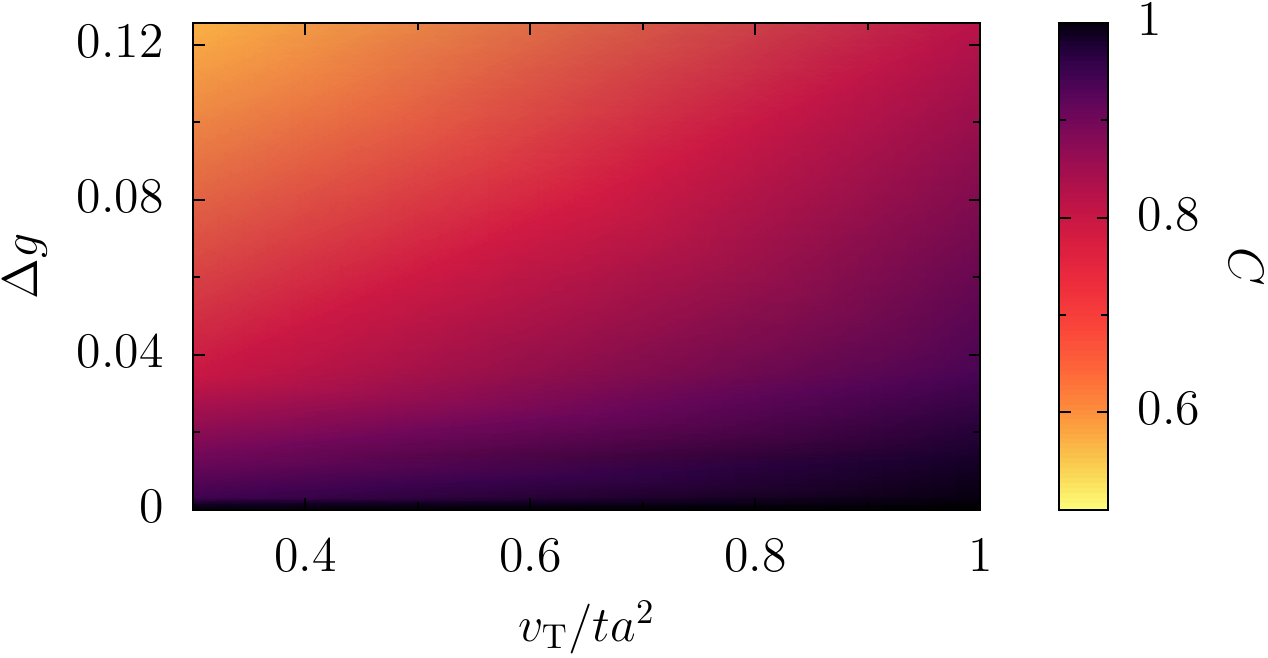}
\caption{\label{figure7} 
Cross-correlation factor \eqref{correlation} as a function of the 
strength $v_\mathrm{T}$ of a local tip (horizontal axis) and the 
deviation from perfect transmission (vertical axis), on the second 
conductance plateau of the QPC in a disordered 2DEG of 
Figs.\ \ref{figure5} and \ref{figure6}.}    	
\end{figure}
\begin{figure}
\includegraphics[width=\linewidth]{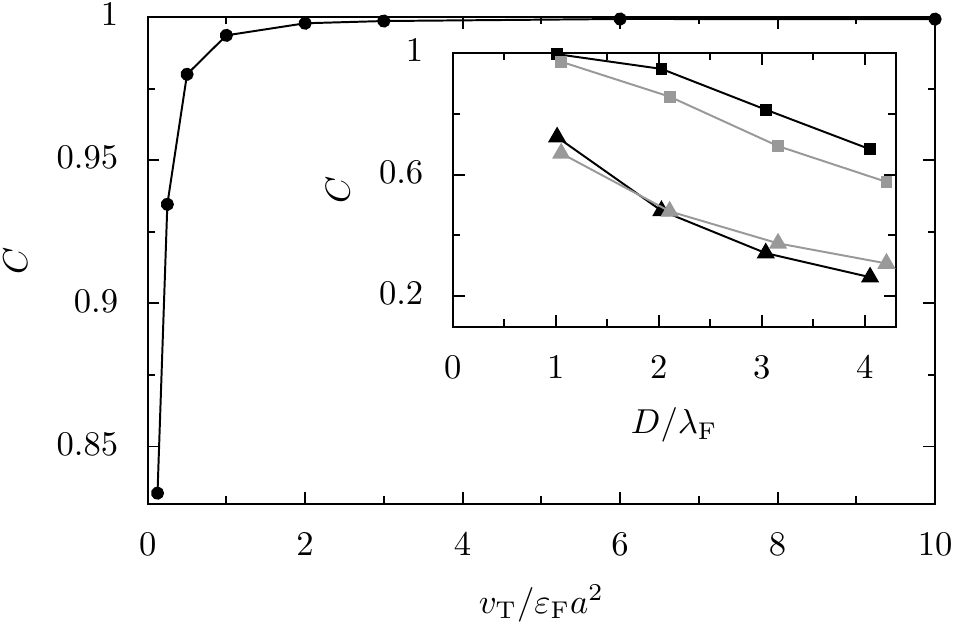}
\caption{\label{figure8} 
Cross-correlation factor $C$ \eqref{correlation}
\textit{vs}.\ the strength of a local tip in the disordered system of 
Fig.\ \ref{figure5}. 
Inset: $C$ \textit{vs}.\ the spatial tip-extension for the smooth 
extended tip shape \eqref{vxy} (gray symbols) and for a 
hard-disk tip (black symbols), in a disorder-free structure. 
Squares and triangles represent the correlation factor between the 
SGM response with the unperturbed PLDOS at the tip center and at the 
classical tuning points, respectively.}    
\end{figure}

 
The results shown in Fig.\ \ref{figure6} demonstrate that the combined operations of extracting the quadratic tip dependence of $\delta g(\br_\mathrm{T})$ and $\lambda_\mathrm{F}/2$-averaging result in nearly perfect extraction of the PLDOS squared.  Still, it is valuable to have a quantitative measure of the quality of this process to answer how well this works as a function of the imperfection of transmission on or near a plateau, and how well it works as a function of tip strength if one chooses just to use $\delta g(\br_\mathrm{T})$ without extracting the quadratic tip-dependence first.  A good measure is given by the cross-correlation factor \cite{kolasinski2013}
\begin{equation}\label{correlation}
  C=\frac{|\overline{(\langle \delta g\rangle -\overline{\delta g })
  (\langle \rho^2\rangle -\overline{\rho^2 })}|}
  {\sigma_{\delta g} \sigma_{\rho^2 }}\, .
\end{equation}
The averages, symbolized by the overlines, are taken over the scanned area in the 
right of the QPC (in contradistinction to the local $O(\lambda_\mathrm{F} /2)$ averages, $\langle ...\rangle$, defined in Sec.\ \ref{sec:numericsg2_local}).  The standard deviations of the two quantities are the usual normalization factors of a properly normalized correlation function.  Applied to $\delta g(\br_\mathrm{T})$ for a range of tip strengths and $\Delta g$'s gives the results shown in Fig.\ \ref{figure7}.  It shows two correlated trends.  
The correlation coefficient decreases with decreasing tip strength and with 
increasing $\Delta g$.   
The value of $v_\mathrm{T}$ for which near perfect correlation is 
achieved depends on the departure $\Delta g$ from perfect transmission.  
Figure \ref{figure8} shows an example for the case of the disordered 
system and tip strengths used in 
Fig.\ \ref{figure5}, where the saturation is reached 
rather quickly as $v_\mathrm{T}/a^2$ increases beyond the Fermi energy. 

Interestingly, the above dependence of $\delta g(\br_\mathrm{T})$ on the tip strength generates a criterion for the validity of perturbation theory.\cite{jalabert2010} 
Note that the criterion for the Born approximation in a one-dimensional scattering problem \cite{Landau} 
$v_\mathrm{T}\ll\EF \lambda_\mathrm{F}$ is consistent with our 
numerical results since the linear extension of the local tip in our tight-binding 
model $a$ is much smaller than $\lambda_\mathrm{F}$. 
In this regime, close to the perfect transmission, 
the second order contribution prevails, and the full SGM response to a local 
tip is highly correlated to the PLDOS squared even for tip strengths larger than 
the Fermi energy. 

\section{Full SGM response for non-local tips}
\label{sec:non-local}

\begin{figure*}
\includegraphics[width=\fwidth]{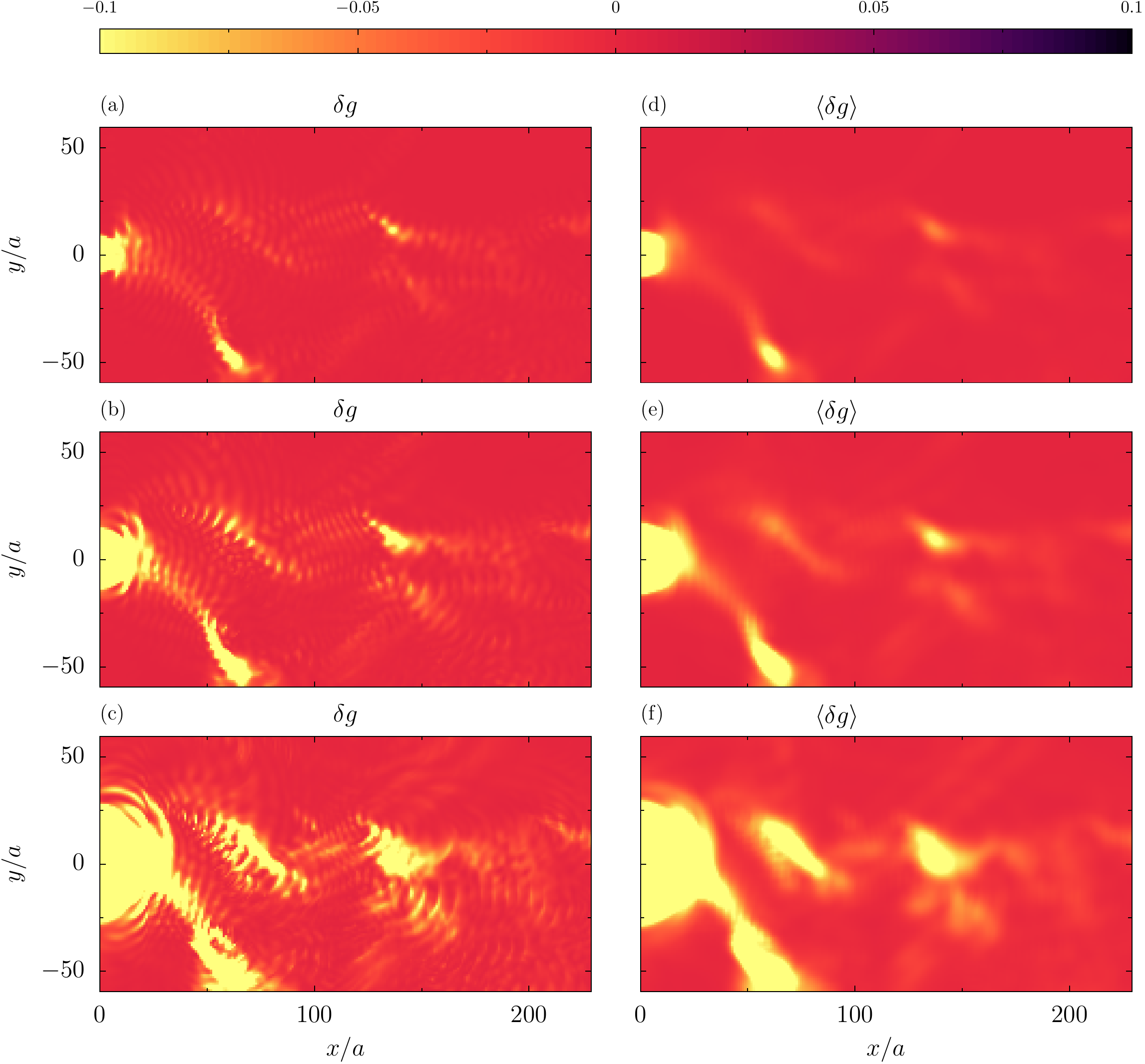}
\caption{\label{figure9}
SGM response calculated using the tip shape \eqref{vxy} for fixed tip potential 
height $v_\mathrm{T}/(2\pi d^2) = 2 \EF$ and varying depletion disk size   
$D=\lambda_\mathrm{F}/2$ (a), $D=\lambda_\mathrm{F}$ (b), and 
$D=2\lambda_\mathrm{F}$ (c). 
Panels (d), (e), and (f) show the averages of the SGM responses 
over a disk of radius $\lambda_\mathrm{F}/2$ for the same tip sizes.}
\end{figure*} 
The case of a local tip, discussed up to this point, is the simplest to analyze, but 
the existing experimental implementations of SGM setups involve extended tips.
Considering the tip as a point charge at a distance $d$ from the 2DEG,
the tip profile in the plane of the 2DEG is of the form
\begin{equation}\label{vxy}
  f(\br)=\frac{1}{2\pi d^2}
  \left[1+\left(\frac{\br-\br_\mathrm{T}}{d}\right)^2\right]^{-3/2} \, .
\end{equation}
Numerical calculations of the electrostatic problem, treating screening within the Thomas--Fermi scheme, result in an approximately Lorentzian (Gaussian) profile when the tip-induced potential does not (does) deplete the 2DEG is \cite{eriksson1996,pala2008,steinacher2015}. Notwithstanding, for tip strengths strong enough to produce depletion, it is observed that the main feature determining the SGM response is the diameter $D$ of the depletion disk, and the details of the tip profile are of lesser importance. Therefore, in our numerical simulations, we adopt the tip profile \eqref{vxy} for all regimes, and express our results in terms of $D=2d [\{v_\mathrm{T}/(2\pi d^2\EF)\}^{2/3}-1]^{1/2}$.

Working in the previously established regime of strong tip strength 
(maximum tip potential $V_\mathrm{T}(\br_\mathrm{T})=v_\mathrm{T}/(2\pi d^2)=2\EF$) 
the SGM response
$\delta g(\br_\mathrm{T})$ for varying tip width $d$ and thus different depletion diameters 
$D$ is present in Fig.\ \ref{figure9}, where the unperturbed conductance and the disorder configuration is the same 
as in Fig.\ \ref{figure5}  (second conductance plateau with 
$\Delta g = 1.3\times 10^{-4}$). 

For $D=\lambda_\mathrm{F}/2$ (panel (a)), the SGM scan resembles that of the 
$\delta$-tip (panel (d) of Fig.\ \ref{figure5}), but with values of 
$\delta g(\br_\mathrm{T})$ 
that are one order of magnitude larger due to the tip extension. 
For larger tip extensions, $D=\lambda_\mathrm{F}$ (panel b) and  
$D=2\lambda_\mathrm{F}$ (panel c), the SGM image gets more blurred and 
some resolution is lost.  This blurring effect is more pronounced 
on the averaged conductance changes, as depicted in the right column panels of 
Fig.\ \ref{figure9}. 

The inset of Fig.\ \ref{figure8} shows the cross-correlation $C$ between 
the non-local SGM and the squared unperturbed PLDOS 
as a function of the depletion diameter $D$. 
Gray symbols correspond to the case of a tip shape of the 
form \eqref{vxy}, the black ones to the case of a hard wall
potential of diameter $D$. The squares represent cross-correlations of the 
SGM response with the PLDOS at the tip center, while triangles depict the 
results obtained when the PLDOS is taken at the classical turning points 
situated at the edge of the depletion disk. Since the classical turning 
point is not determined uniquely in the presence of disorder, the data in 
this inset are for the disorder-free structure. We have checked that 
including disorder does not change significantly the results when the 
tip center is taken as the reference point for the PLDOS. 
For both tip shapes, \eqref{vxy} and hard wall, and independent of where 
the PLDOS is taken, the cross-correlation decreases with increasing 
depletion diameter $D$.
 
If the PLDOS is taken at the classical turning point (triangles)
instead of the tip center (squares), the SGM response becomes 
less correlated with the PLDOS. The classical argument 
of Ref.\ \onlinecite{heller2003} that predicts that a large circular 
hard-wall tip does image the local properties of the unperturbed 
structures by reflecting back the classical trajectories that hit 
the tip with normal incidence does not appear as a limiting case of 
our results. One reason could be that our numerics did not reach 
sufficiently large depletion disks with $D\gg\lambda_\mathrm{F}$ 
to observe such a behavior. \cite{jalabert2015}
Another reason could be that the SGM 
response in the classical limit is not well correlated with the 
squared PLDOS as in the case of local tips, though another link 
to the PLDOS at the classical turning point of a large disk cannot 
be excluded from our study.

\section{Conclusions}
\label{sec:conclusions}

With regards to the quest of extracting information about local 
electronic properties in phase-coherent devices from SGM measurements, 
we have investigated the correspondence between the SGM response in the 
vicinity of a QPC and the unperturbed PLDOS. Only on the first conductance step could the PLDOS be shown to settle an upper bound for the magnitude of the first-order SGM correction. 
We have shown analytically that the unperturbed PLDOS squared is 
unambiguously related to the second-order conductance correction 
induced by a local tip, provided that the system is time-reversal 
symmetric and the QPC is tuned to perfect transmission. 
The second-order correction dominates the SGM response on a 
``perfect'' conductance plateau if the tip strength is not too strong.
If the QPC transmission is imperfect, the exact correspondence is 
broken, and the departures are quantified with a perturbation theory. 
It does not depend on fine details of the setup, but rather on the 
scale of the unperturbed conductance's deviation from perfection, 
$\Delta g$.

We have demonstrated that a correspondence between the locally 
averaged second-order SGM response and the PLDOS survives for 
imperfect transmission obtained when the highest propagating eigenchannel is not completely open. 
Numerical simulations within a recursive Green function approach have 
confirmed our analytical findings and shown that they also hold in the 
case of disordered systems.  

Moreover, we found that in the case of a local tip, and sufficiently 
small $\Delta g$, the full SGM response is related to the PLDOS once 
the tip is strong enough such that the second-order conductance 
correction dominates.  

In the case of non-local tips, where the depletion disk created by 
the tip exceeds half the Fermi wavelength, the correspondence between 
the SGM response and the PLDOS established for weak local tips degrades 
with increasing depletion disk radius. 

Most SGM experiments are performed in high mobility 2DEGs in which 
the Fermi wavelength is smaller than the depletion disk under the tip.
In that case the relationship between the SGM response and the PLDOS 
squared degrades and beyond a large enough radius cannot be used 
directly to and unambiguously extract local electronic properties.
For experiments in the weakly invasive regime, the resolution of 
the SGM response is also limited by the width of the tip 
potential.\cite{steinacher2017} 
One way to approach the regime where the direct link is valid would be 
to use systems with lower Fermi energy and thus larger Fermi wavelength.

In a very recent SGM experiment \cite{haeusler2017} performed using ultracold 
atom gases, a tightly focused laser beam played the role of the tip and could be  
scanned in the neighborhood of a QPC attached to two atom reservoirs. In this case 
a resolution better than \unit[10]{nm} with a tip size well 
below $\lambda_\mathrm{F}$ was obtained. In this regime, we expect that the 
relationship established between the SGM response and the LDOS is applicable. 

\acknowledgments
The authors are grateful to B.\ Braem, B.\ Brun, K.\ Ensslin, 
C.\ Gold, C.\ Gorini, T.\ Ihn, C.\ P\"oltl, 
R.\ Steinacher, and G.\ Weick for useful discussions.  
Financial support from the French National Research
Agency ANR through Projects No.\ ANR-14-CE36-0007-01 (SGM-Bal) 
and No.\ ANR-11-LABX-0058\_NIE (Labex NIE within ANR-10-IDEX-0002-02)
is gratefully acknowledged.

\appendix

\section{Scattering wave-functions}
\label{sec:AppA}
In this Appendix we recall the main concepts of scattering theory for quantum 
transport in view of the application to the SGM setups implemented through the text.
The incoming lead modes $\varphi_{1(2),\varepsilon,a}^{(-)}(\br)$  are given by 
\begin{subequations}
	\label{leadmodes} 
	\begin{align}
	\varphi_{1,\varepsilon,a}^{(-)}(\br) & =  
	\frac{c}{\sqrt{k_{a}}} \ \exp{[ i k_{a}^- x]}
	\ \phi_{a}(y) \ , \quad x <  \ 0  \\
	\varphi_{2,\varepsilon,a}^{(-)}(\br) &=  
	\frac{c}{\sqrt{k_{a}}} \ \exp{[- i k_{a}^- x]} 
	\ \phi_{a}(y) \ , \quad x >  \ 0 \, ,
	\end{align}
\end{subequations} 
where $\phi_{a}(y)$ is the wave-function of the $a$\textsuperscript{th} 
transverse channel along the lead and $k_{a}^-$ 
the longitudinal wave number $k_a$ with an infinitesimal imaginary part 
necessary for incoming modes.
We note $c=[m^*/(2\pi\hbar^2)]^{1/2}$, with $m^*$ the effective electron mass.
In writing $x<0$ and $x>0$ we mean the asymptotic condition in the left and 
right leads, respectively (see Fig.\ \ref{figure1}).

In the basis of the $2N$ incoming modes the scattering matrix is defined by
\begin{equation}
\label{S}
S = \left( \begin{array}{cc}
r & t' \\
t & r'
\end{array} \right) \, .
\end{equation} 
The incoming modes give rise to outgoing scattering states, which in the asymptotic regions can be expressed as,
\begin{subequations}\label{allscats}
\begin{align}
\Psi_{1,\varepsilon,a}(\br) 
&= 
\left\lbrace 
\begin{array}{ll}
\varphi_{1,\varepsilon,a}^{(-)}(\br) + \sum_{b=1}^{N} r_{ba} \, \varphi_{1,\varepsilon,b}^{(+)}(\br),
 & x < 0 \\
\sum_{b=1}^{N} t_{ba} \, \varphi_{2,\varepsilon,b}^{(+)}(\br), & x > 0  
\end{array} 
\right. 
\\
\Psi_{2,\varepsilon,a}(\br) 
&= 
\left\lbrace 
\begin{array}{ll}
\varphi_{2,\varepsilon,a}^{(-)}(\br) + \sum_{b=1}^{N} r^{\prime}_{ba} \, 
\varphi_{2,\varepsilon,b}^{(+)}(\br), & x > 0 \\
\sum_{b=1}^{N} t^{\prime}_{ba} \, \varphi_{1,\varepsilon,b}^{(+)}(\br), & x  < 0 
\end{array} 
\right. 
\end{align}
\end{subequations}
%
in terms of the matrix elements of the reflection $r$ ($r'$) and transmission $t$ ($t'$) submatrices for electrons impinging from the left (right) lead.

The scattering matrix is conveniently expressed in the polar decomposition \cite{mello2004}, 
which in the case of time-reversal symmetry takes the form
\begin{equation}
S = \left( \begin{array}{cc}
u_{1}^\mathrm{T}		&  0	\\
0		&  u_{2}^\mathrm{T}
\end{array} \right)
\left( \begin{array}{cc}
-{\cal R}	& \hspace{0.5cm} {\cal T}	\\
{\cal T}	& \hspace{0.5cm} {\cal R}
\end{array} \right)
\left( \begin{array}{cc}
u_{1}	&  0	\\
0	&  u_{2}
\end{array} \right) \, .
\label{polarS}
\end{equation}
${\cal R}$ and ${\cal T}$ are diagonal reflexion and transmission submatrices, 
while $u_1$ and $u_2$ are unitary matrices. The transmission eigenmodes 
take the form \cite{gorini2013}
\begin{subequations}
	\label{Teigenmodes} 
	\begin{align}
		\varrho_{1,\varepsilon,m}^{(-)}(\br) &=
		\sum_{a=1}^{N} \left[u_{1}\right]_{m a}^{*} \
		\varphi_{1,\varepsilon,a}^{(-)}(\br)
		\, , \quad x <  \ 0 \ ,
		\\
		\varrho_{2,\varepsilon,m}^{(-)}(\br) &=
		\sum_{a=1}^{N} \left[u_{2}\right]_{m a}^{*} \
		\varphi_{2,\varepsilon,a}^{(-)}(\br)
		\, , \quad x >  \ 0 \ .
	\end{align}
\end{subequations}

Identifying \eqref{S} and \eqref{polarS}, the transmission and reflexion 
submatrices can be expressed as 
$t = u_{2}^{T}{\cal T}u_{1}$, $t^{\prime } = u_{1}^{T}{\cal T}u_{2}$, 
$r =- u_{1}^{T}{\cal R}u_{1}$, and $r' = u_{2}^{T}{\cal R}u_{2}$. 
Thus, $t^\dagger t = u_{1}^{\dagger}{\cal T}^2u_{1}$ and 
$t^{\prime \dagger} t^{\prime} = u_{2}^{\dagger}{\cal T}^2u_{2}$. 
  
Considering the vector of coefficients 
$C_{1(2)m}=([u_{1(2)}^*]_{m1},[u_{1(2)}^*]_{m2}, ...)^T$ of the 
transmission eigenmode $\varrho_{1,\varepsilon,m}^{(-)}(\br)$, one can write
\begin{equation}
  t^{\dagger} t C_{1m}=u_{1}^{\dagger}{\cal T}^2u_{1}C_{1m}
  ={\cal T}_m^2C_{1m} \, .
\end{equation}
The second equality stems from the definition of $C_{1m}$ and implies
that $C_{1m}$ is an eigenvector of $t^{\dagger} t$ with the eigenvalue 
${\cal T}_m^2$. In the same way, one finds that $C_{2m}$ is an eigenvector 
of  $t'^{\dagger} t'$ with the same eigenvalue. 
  
The scattering eigenstates in the region $x>0$ for an incoming transmission 
eigenmode $\varrho_{1,\varepsilon,m}^{(-)}(\br)$ are obtained as
$t C_{1m}=u_2^T{\cal{T}}u_1C_{1m}$. Using again the definition of $C_{1(2)m}$
and the unitarity of $u_1$ we find
\begin{equation}
   t C_{1m}= {\cal {T}}_{m}C_{2m}^* \, ,
\end{equation} 
and similarly 
\begin{equation}
  r C_{1(2)m}=\mp {\cal {R}}_{m}C_{1(2)m}^* \, .
\end{equation} 
Thus, the basis of scattering eigenfunctions is asymptotically given by
\begin{subequations}\label{xis}
		\begin{align}
			\chi_{1,\varepsilon,m}(\br) 
			&= 
			\left\lbrace 
			\begin{array}{ll}
				\varrho_{1,\varepsilon,m}^{(-)}(\br) 
				- {\cal R}_m \, \varrho_{1,\varepsilon,m}^{(+)}(\br)\, ,
				&   x <  0  \\
				{\cal T}_m\, \varrho_{2,\varepsilon,m}^{(+)}(\br)\, , &  x > 0   
			\end{array} 
			\right. \label{scatst1} 
			\\
			\chi_{2,\varepsilon,m}(\br) 
			&= 
			\left\lbrace 
			\begin{array}{ll}
			{\cal T}_m \, \varrho_{1,\varepsilon,m}^{(+)}(\br)\, , &  x < 0 \\
			\varrho_{2,\varepsilon,m}^{(-)}(\br) +  {\cal R}_m \, 
			\varrho_{2,\varepsilon,m}^{(+)}(\br)\, , &   x >  0  
			\end{array} 
			\right.\, . \label{scatst2}
		\end{align}
	\end{subequations}
The PLDOS \eqref{ldos}, as well as the conductance corrections \eqref{fullg1_t_basis} and \eqref{fullg2_t_basis} 
are conveniently discussed when expressed in the basis of scattering eigenfunctions.  

\section{Hilbert transform of LDOS}
\label{sec:AppB}

In this Appendix the term of Eq.\ \eqref{fullg2_t_basis} 
containing the principal part is related with the LDOS.

For a $\delta$-tip, we can write
\begin{widetext}
\begin{equation}
\mathrm{Tr} \left[ \mathcal{R}\mathcal{T} 
\mathcal{U}^{2 \bl}(\EF,\bepsilon)\mathcal{U}^{\bl 1}(\bepsilon,\EF) \right] 
= \frac{\rho_{\bl \bepsilon}(\br_\mathrm{T})}{2\pi}
 \mathrm{Tr}\left[\mathcal{R} \mathcal{T} \mathcal{U}^{2 1}(\EF,\EF)\right]\, ,
\end{equation}
and therefore
\begin{equation}\label{eq:hilbert}
\frac{-4\pi}{v_\mathrm{T}^2}\sum_{l=1}^2 
\mathcal{P}\int_{\varepsilon_1^{(\mathrm{t})}}^\infty 
\frac{\dif\bepsilon}{\bepsilon-\EF}
\mathrm{Im}\left\{\mathrm{Tr} \left[ \mathcal{R}\mathcal{T} 
\mathcal{U}^{2 \bl}(\EF,\bepsilon)\mathcal{U}^{\bl 1}(\bepsilon,\EF) \right] \right\}
= \frac{g^{(1)}(\br_\mathrm{T})}{2}
\left\{\frac{1}{\pi} \mathcal{P}\int_{\varepsilon_1^{(\mathrm{t})}}^\infty 
\frac{\dif\bepsilon }{\EF-\bepsilon } \rho_{\bepsilon} (\br_\mathrm{T})
\right\} \, .
\end{equation}
\end{widetext}
Since the LDOS vanishes for $\bepsilon<\varepsilon_1^{(\mathrm{t})}$, 
the lower limit of the integral can be taken as $-\infty$. Given that 
the LDOS is proportional to the imaginary part of the diagonal 
Green function $\mathcal{G}_{\varepsilon}(\br,\br)$, the curly 
bracket at the right-hand-side represents a Hilbert transform (with respect to the energy variable) leading to the real part of 
$\mathcal{G}_{\varepsilon}(\br,\br)$, as indicated in the discussion 
following Eq.\ \eqref{eq:eta}.  
The term \eqref{eq:hilbert} contributing to $g^{(2)}$ and fulfilling 
a Kramers-Kronig relation with the LDOS, is dominated by the 
contribution of the latter close to the Fermi energy.

The emergence of the Hilbert transform of the LDOS has been signaled 
for the first-order SGM correction of a one-dimensional 
system. \cite{pala2008}
In our case it appears in the contribution \eqref{eq:hilbert} 
to the second-order correction \eqref{fullg2_t_basis}, and it is 
not restricted to a one-dimensional setup. Such a contribution, 
also proportional to the first-order correction 
$g^{(1)}(\br_\mathrm{T})$, is necessarily very small when the QPC 
operates close to the condition of conductance quantization. 

\bibliography{references} 

\end{document}